\documentstyle[12pt, fleqn, amsfonts]{article}
\begin{document}
\begin{titlepage}
\title{On Ordering  of Operators in Canonical Quantization \\
 in Curved Space}
\thanks{quant-ph/0101016}
\author {\large E.A.Tagirov  \\
N.N.Bogoliubov Laboratory of Theoretical Physics,\\
Joint Institute for Nuclear Research, Dubna, 141980, Russia; \\
e-mail: tagirov@thsun1.jinr.ru}
\date{}
\end{titlepage}
\maketitle

\begin{abstract}
Ambiguities arising in different approaches (canonical, quasiclassical,
path integrals) to quantization  are discussed by an
example  of the mechanics of a point-like
particle in the Riemannian space. A way to select
a single rule of quantization is proposed by  requiring
of consistency of  the quantum mechanics' following from
the canonical and quasiclassical approaches. This rule selects  also
a unique  definition of the path integration.  A geometric
interpretation
of noncovariance of the canonical  Hamilton operator with respect to
diffeomorphisms of the configuration space  is proposed.
\end{abstract}

\newcommand{\qarr}{\stackrel{\cal Q}{\longrightarrow}}
\newcommand{\im}{{\rm i}}
\newcommand{\Sg}{\Sigma}
\newcommand{\bgt}{\bigotimes}
\newcommand{\ptl}{\partial}
\newcommand{\Sche}{ Schr\"odinger equation\ }
\newcommand{\Schr}{Schr\"odinger representation\ }
\newcommand{\ov}{\overline}
\newcommand{\std}{\stackrel{def}{=}}
\newcommand{\stc}{\stackrel}
\newcommand{\h}{\hbar}
\newcommand{\iq}{q^{(i)}}
\newcommand{\ip}{p_{(i)}}
\newcommand{\hip}{\hat p_{(i)}}
\newcommand{\hiq}{\hat p_{(i)}}
\newcommand{\beq}{\begin{equation}}
\newcommand{\nde}{\end{equation}}
\newcommand{\bea}{\begin{eqnarray}}
\newcommand{\nda}{\end{eqnarray}}
\newpage
\section {Introduction}

Various  procedures of quantization (canonical, path integration,
quasiclassical,  deformational) of   classical mechanics  are
more or less well--developed mathematical  theories.  However,
being used to construct the  quantum mechanics (QM)\footnote
{I will use the  term `quantum mechanics' for the
restricted problem  of quantization of the minimal set of
observables that
is  necessary to describe the dynamics of a physical system,
while `quantization' will mean a more general  problem  of  mapping
from  a sublalgebra of the Poisson algebra of functions on a
phase space onto a set of operators on a Hilbert space, or
for modifications of this problem.} for a concrete classical system,
 the  procedures occur to be ambiguous. For example, the result of
canonical quantization  depends generally  on  the definition of
ordering
of operators of observables in their products, see, e.g.,
\cite{BER}, though there is no ambiguity for the standard
oscillator--like systems. This ambiguity   manifests itself also
in the path integral formalism, see, e.g., \cite{OLIV}.
This situation  does not trouble apparently  the mathematicians,
because they consider different versions of the same approach,
 as  cohomologically equivalent \cite{FL}.
However,  the ambiguities are essential in physics, because they
can lead, for example, to different Hamilton operators and, thus,
are  unequivalent from a physicist's point of view.

A point-like, chargless and spinless particle moving along a geodesic
of the $n$--dimensional Riemannian  space $V_n$  is an especially
interesting example of a system,  to which different QMs correspond
under canonical
quantization. This  is the simplest and, therefore,  elementary physical

object,  and its dynamics has a clear geometrical meaning.  At the same
time,
there are a lot of  more complicated conceptual systems whose dynamics
can be modelled as a motion in some $V_n $ (see, e.g. \cite{GRO}).

In the present paper, it is shown that the ambiguity of ordering
of operators can be removed  if the condition of consistency  (to an
extent
that  can be reached)  between the canonical and quasiclassical
approaches  to QM of the elementary particle  in $V_n$ is imposed.
At  first  sight, it looks  strange to deduce a general rule of ordering

from the consideration  of a concrete system, and,  moreover,
from the consideration of a
particular operator of an observable, namely, the Hamilton one.
However, if the rule  is supposed to be  universal,
and  only a  single  version is
provided by the condition  of the consistency in the particular but
fundamental and geometrically determined case, then it is natural
to  accept the  selected rule as the general one.  After that,
it  selects "the physically true" definition of the  path integral.

The paper is organized as follows. In Sec.2, the form
of the Hamilton operator  originally obtained by B.DeWitt \cite{DW}
from the  quasiclassical (WKB)  propagator is discussed.
The general scheme of canonical
quantization and  its  particular versions determined by
the Weyl and Rivier orderings are recalled in Sec.3.
In Sec.4, the Hamilton operators in QM in $V_n$  for the mentioned
orderings are written out and compared with the DeWitt's one; it is
shown
here that the consistency can be achieved
only for a single combination of the Weyl and Rivier orderings.
This combination  is proposed as a distiguished rule of
canonical quantization in general. Relation of the selection of the rule
of
ordering to definition
of the  path integral for propagator is discussed  in Sec.5.
Short conclusions are made in Sec.6.

\section {Quasiclassical quantum potential}

Let $\omega_{ij}(\xi), \quad i, j, k,...= 1, ..., n, \quad
\{\xi^i\} \in V_n $, is the metric tensor of $V_n$ ,
the latter being supposed an elementary manifold.
Then  the representation space (the space of states)
can be taken as $L^2 (V_n ; {\Bbb C}; \sqrt{\omega}\ d^n \xi) $.
The  general form  of  a one-particle
Hamilton operator in the \Schr is the same  for the canonical and
quasiclassical approaches\footnote{The index 0 is attached to
  $\hat H_0$ for consistency with notation in \cite{TAG1, TAG2}
where it denotes a nonrelativistic Hamilton operator. In   the
present context, relativistic and nonrelativistic classical
theories are equivalent and the corresponding quantum ones can
differ (This is
one more  problem of quantization.) Thus, QM is considered
here on a  nonrelativistic level, see also \cite{TAG2}}:
\beq
{\hat H_0} \ = \
\Delta (\xi)+ \ V_q (\xi) \cdot {\bf \hat 1},
\quad \ \xi\in V_n, \label{HAM1}
\nde
where $\Delta (\xi) $ is the Laplace--Beltrami operator in
$V_n$ and $V_q (\xi)$ is the so called quantum potential.

In the quasiclassical approach, DeWitt \cite{DW} came to
the Hamilton operator (\ref{HAM1})  with
\beq
 V^{\rm{(DW)}}_q (\xi) = -  \frac{\h^2}{2m}\ \frac16 R(\xi), \label{V1}
\nde
  instead of the expected
$V_q (\xi) = 0$; here $R(\xi)$ is the scalar curvature .

DeWitt \cite{DW}  started with
the following conjecture  on a one-particle propagator\footnote{
Here, I change slightly the notation in eq.(7.8) in \cite{DW}  and
restrict
the consideration to the time--independent metric tensor
$\omega_{ij} (\xi)$}:
\beq
<\xi',t'| \xi, t> =
\omega^{-1/4} (\xi') D^{1/2} (\xi'|\xi)\omega^{-1/4} (\xi)\
\exp\left(-\frac\im\h S(\xi',t'|\xi, t)\right) \label{PROP}
\nde
where $D$ is the Van Vleck determinant, and $S(\xi',t'|\xi, t) $
is a solution of the Hamilton--Jacobi equation in  both the
sets of arguments. This propagator
is a generalization  to $V_n$  of the WKB-propagator
constructed by Pauli \cite{PAU}  for a particle in an
electromagnetic field in the flat space. Considering the limit
$t' \rightarrow t \ (\xi'\rightarrow \xi) $ {\it along the geodesic line

connecting $\xi'$ and $\xi$}, DeWitt comes to the following \Sche
\beq
\im \h \frac\ptl{\ptl t'}<\xi'| \xi> + \ \frac{\h^2}{2m} \left(\Delta
(\xi')
+ \frac16 R (\xi')\right) <\xi'| \xi>  = o (\xi' - \xi)) <\xi'| \xi>
\label{DW2}
\nde
and, thus, to eq.(\ref{V1}).

This was an outstanding result that  surprised DeWitt himself, because
it was for the first time when the curvature appeared explicitly
in a fundamental equation,  thus breaking
the (weak) Principle of Equivalence. (A clear exposition of the
Principle  can be found in   \cite{WEIN}). Since beginnings
of QM  there was a confidence that $V_q (\xi)= 0$,
based on the work by Podolski  \cite{POD} on QM in the curvilinear
coordinates in the Euclidean space.  Much later, Sniatycki
\cite{SN} came to the same result   using
the Blattner-Kostant-Sternberg kernel in the framework of geometric
quantization.  At last, for $n=3$,  there is  a support in favor of
$V^{\rm{(DW)}}_q (\xi)$ from the relativistic theory of  scalar
field: for the space-time
$ T \bigotimes V_3$ eq.(\ref{HAM1}),  is  a  non-relativistic
consequence  of the  quantum theory of a scalar field
coupled conformally to the geometry, (see details in \cite{TAG1}).

Consider now the  general scheme of the canonical formalizm.

\section {Canonical quantization in $V_n$}

We start with  the phase space  $X_{2n} = R^* \bigotimes V_n $,
the trivial cotangent  bundle  over the  configurational space
$V_n$, in  which the Darboux (canonically conjugate) coordinates
\beq
\{p_i \}\in\ R^*_n,  \quad \{ \xi^i\}  \in V_n
\nde
are introduced. We assume also that all coordinate lines for  $
\{\xi^i\}$
are complete and not closed, that is $-\infty < \xi^i < \infty $.
In this sense, $ \xi^i\}$  are similar  to  the Cartesian coordinates;
this assumption is necessary because,  in the canonical quantization,
one starts usually with polynomials as observables and
genereralize the class of functions by use of the Fourier transforms
\cite{BER}. A physical interpretaion  of this very restrictive
assumption is that local manifestations of the space curvature are
taken into account.

 For our purposes, {\it quantization} (see, e.g., \cite{KIR})
can be defined as a map
\beq
{\cal Q}: \ {\cal F}_{2n} \ni f(p, \xi) \longrightarrow \hat f\
(\mbox{an operator acting  in\ } {\cal H} \equiv
L^2 (V_n ;\ {\Bbb C};\ \sqrt{\omega}\ d^n\xi)),
\nde
where ${\cal F}_{2n}$ is an appropriate subalgebra of the Poisson
algebra of  $C^\infty(X_{2n})$--functions;  the map {\cal Q} is assumed
to
satisfy the following  postulates of quantization:\\

1) $ \quad  1\ \rightarrow {\bf\hat 1} $;

2) $\quad  \{f , g\}\ \rightarrow \
i\h^{-1}   (\hat f \hat g - \hat g \hat f)
\std \im\h^{-1} [\hat f, \hat g]$ \quad
where $\{.,.\}$ is the Poisson bracket  on $X_{2n}$;

3)$\quad    \ \hat{\ov f} = (\hat f)^\dagger $ where the  overline and
the dagger denote, respectively, the  complex conjugation and the
Hermitean
conjugation  with respect to the inner product in $\cal H$ ;

4)\quad   for a complete set of functions
$f^1 (x) , ..., f^n (x), \ f^i \in {\cal F}(X_{2n}) ,
\quad (\{x\} \sim \{p, \xi \}) \in X_{2n}) $,
the corresponding operators  $\hat f^1, ..., \hat f^n$  also form
a complete set,  that is,  if
 $\left[\hat f,\ \hat f^i \right] = 0  $ for any  $ i = 1,...,n  $, then

$\hat f\,=\, \hat f (\hat f^1, ..., \hat f^n) $.  \\

The real functions $ f(x) \in {\cal F}(X_{2n})$  are observables
for a classical system on $X_{2n}$. The Darboux coordinates are
particular cases of such functions, and, since any $f(x)$ can be
expressed
through these coordinates, the latter can be called basic classical
observables. The corresponding  quantum operators of
the observables  acting in $L^2 (V_n;\ {\Bbb C};\ \sqrt{\omega}\ d^n
\xi)$
are \cite{DW}
\beq
      \xi^i\ \qarr\ \hat\xi^i = \xi^i \cdot\hat {\bf 1},\quad
     p_j\ \qarr\  \hat p_j  = - \im \h(\ptl_j +
\frac14 \ptl_j \ln \omega)              \label{QP}
\nde

The main problem  is further  to construct the map
$f(p, \xi)\ \qarr\ \hat f $  in terms of the basic observables
$\hat p, \ \hat \xi^i$.  In the canonical approach, one proceeds
as follows \cite{BER}: \\
\noindent
1) Starts  with (real)  polynomial observables:
\beq
f(p, \xi) = f + \sum_{a,b} \ f_{i_1...i_k}^{j_1...j_l}\
 \xi^{i_1} ... \xi^{i_k}\ p_{j_1} ... p_{j_l};  \label{POL}
\nde
2) substitutes $p_j,\ \xi^i $  by  $\hat p_j,\ \hat\xi^i $,
respectively;\\
3) hermitizes the obtained operator by a certain rule of  ordering
of $\hat p_j,\  \hat\xi^i$;\\
4) generalizes  the rule, if possible, to a  class
of observables $f (p, \xi)$ wider, than that of polynomials.\\

There are infinitely many realizations of this scheme. They are
classified, e.g., in \cite{AGWO}. Consider now two examples, one
of which, the Weyl ordering, is the most popular one,  and  the both
of them are important for our further consideration.\\

{\it Example 1.} According to \cite{BER},  the Weyl ordering as
applied to  polynomial
(\ref{POL}) consists in complete symmetrization of operators  $\hat p$
and
$\hat q$ in each monomial in the right-hand side of eq.(\ref{POL})
after substitutions
$p_i \rightarrow \hat p_i,\  q^j \rightarrow \hat q^i $.
The result can be represented  in $V_n$  as
\bea
f(p, \xi) & &\qarr\   (\hat f \psi) (\xi) \nonumber\\
&=& (2 \pi \h)^{-n} \omega^{-\frac14} (\xi)\,
\int d^n\tilde\xi\,  \, d^n p\
\exp\left(-\frac{\im}{\h} (\xi^i - {\tilde \xi}^i) p_i \right)
f\left(p, \frac{\xi +\tilde \xi}{2}\right)\
\omega^{\frac14}(\tilde\xi)\,  \psi(\tilde \xi). \label{WEY}
\nda
Further, this result is taken as a rule determining an operator
$\hat f$ for any function $f(p, \xi)$ for which the expression
has a meaning.\\

{\it Example 2.}  Rivier \cite{RIV} proposed
a correspondence  that is equivalent to  the following  one
\bea
& &f(p, \xi)\qarr\ (\hat f \psi) (\xi) \nonumber\\
& & = (2 \pi \h)^{-n}
\omega^{-\frac14} (\xi)  \int d^n\tilde\xi\,  d^n p\
\exp\left(-\frac{\im}{\h} (\xi^i - {\tilde \xi}^i)p_i\right)\
\frac{f(p, \xi)
+ f(p, \tilde \xi)}{2} \  \omega^{\frac14}(\tilde\xi) \, \psi(\tilde
\xi).  \label{RIV}
\nda
An  argument in favor of this rule  was that it provides
 one-to-one correspondence between
 infinitesimal canonical transformations on $ X_{2n}$ and
infinitesimal unitary transformations in  $\cal H$.
It is easily shown  that the
Rivier rule  of quantization (\ref{RIV}) leads to the  following
ordering  of each term in (\ref{POL}): .
$$ \frac12 (\hat\xi^{i_1} ... \hat\xi^{i_k}\, \hat p_{j_1} ...\hat
p_{j_l} +
\hat p_{j_1} ...\hat p_{j_l}\, \hat\xi^{i_1} ... \hat\xi^{i_k}) $$

 A  question of fundamental importance is: which rule of ordering
should be used to obtain  QM of the most  elementary   physical
system, a point-like, spinless and chargeless particle, if external
gravitation is switched on in the form of nontrivial
time-independent space metric?

\section{Quantum Mechanics of Geodesic Motion in $V_n$}

Having the correspondence (\ref{QP}) and reducing the general problem
of quantization to a formulation of QM corresponding to the geodesic
motion of a particle ,  one should  only  construct a quantum
counterpart
for the Hamilton function
\beq
  H_0 = \frac1{2m}\ \omega^{ij} (\xi)\ p_i p_j
\nde
where $m $  is the mass of the particle
According to the scheme given  above, the problem  consists  in the
choice of  ordering of operators
$$\hat\xi^i,\ \hat p_j,\ \  {\hat\omega}^{ij} \std \omega^{ij}(\hat\xi)
=
\omega^{ij}(\xi)\cdot\hat {\bf 1}.
$$
{\it Note that the metric tensor becomes now  an operator} determined
in fact by the von Neumann  rule  for quantization \cite{NEU}:
if   $ f \stackrel{\cal Q}{\longrightarrow} \hat f $, then
$ g(f) \stackrel{\cal Q}{\longrightarrow} g(\hat f)$.
Then, the variety of natural orderings which  give
$\hat f = \hat f^\dagger $ for $ f(p, \xi) =\ov f(p, \xi)$,
 is exhausted by linear  combinations
of the  Weyl and Rivier orderings considered  above. However,
other  more exotic versions of ordering might be introduced by
representing  $\omega^{ij}$  as a product of  other observables
and taking their symmetrizations with $\hat p_i$ and $\hat p_j$,
as it is done in \cite{GRO2}, but we do not consider
here these possibilities.

The Hamilton  operators  for the Weyl and Rivier orderings
are,  respectively,
\bea
  \hat H^{\rm{(Weyl)}}_0 & = & \frac1{8m} (\hat p_i\hat p_j\
\omega^{ij}(\hat \xi)
+ 2\ \hat p_i\ \omega^{ij}(\hat \xi)\ \hat p_j
+ \omega^{ij} (\hat \xi) \ \hat p_i\hat p_j )\\
  \hat H^{\rm{(Riv)}}_0 & = & \frac1{4m} (\hat p_i\hat p_j\
\omega^{ij}(\hat \xi)
+  \omega^{ij} (\hat \xi)\ \hat p_i\hat p_j )
\nda
Using representation (\ref{QP})  for the basic operators, after some
algebra, one
obtains   both the Hamilton operators in the form of eq.(\ref{HAM1})
with a  quantum  potential, respectively,
\bea
\hat H^{\rm{(Weyl)}}_0 & = & -\frac{\h^2}{2m} \left(\Delta(\xi)
+\frac12 \ptl_j(\omega^{ij}\gamma_i)
+\frac14\ptl_i\ptl_j \omega^{ij}
+ \frac14\omega^{ij}\gamma_i \gamma_j\right) \label{HWEY}  \\
  \hat H^{\rm{(Riv)}}_0  & = & -\frac{\h^2}{2m}\left(\Delta(\xi)
+\frac12\ptl_j(\omega^{ij}\gamma_i)
+ \frac14\omega^{ij}\gamma_i \gamma_j \right)  \label{HRIV} \\
\nda
where $\gamma_i \std {\gamma^k}_{ki}$, and ${\gamma^k}_{ij}$ are the
Christoffel symbols for metric tensor $\omega_{ij}$ .

One sees that  the quantum potentials in  both the cases   are
noncovariant,
with respect to diffeomorphisms of $V_n$, though the "kinematic"
term  $-(\h^2 /2m) \Delta (\xi)$ is a scalar
operator. It can be explained by that the coordinates $\xi^i$
play in the theory  a two-fold  role: they provide  $V_n$ by a manifold
structure
and, at the same time,  as a part of  $2n$ coordinates
of the phase space $X_2n$, they are classical observables.
Correspondingly,  $\hat \xi^i$ form a complete set of quantum
observables,
in terms of which the preparation and observation
of a state of the system is performed.

The dependence of  the quantum  dynamics on a  choice of observables
(that form "the quantum statics" according to  \cite{SAD}) via the
Hamilton operator is  hidden in
the standard QM  in the flat space  by   use of  Cartesian coordinates
as position observables, for which $V_q = 0$.  There, a use of
curvilinear cordinates, e.g.,
the spherical   ones,  is done {\it a posteriori}, after
introduction of the Hamilton  operator in  terms of  basic observables
which are firmly related to  Cartesian coordinates.
The curvilinear coordinates are  used only as an auxiliary mathematical
tool and, thus, play only the first part of  the two-fold role
mentioned above. The situation changes
drastically in $V_n$ where one forced to use  curvilinear coordinates as

basic observables from the begining, and  the  problem of
noncovariance of the quantum potential arises inevitably.

This problem is often circimvented  by taking
$\hat H_0  =  -\frac{\h^2}{2m} \Delta(\xi)$  as a postulate.
However, it leads us out the framework of the canonical formalism, and,
what is much worse,  one comes then to noncovariance of the exponent
term in the path integral, see \cite{OLIV} and Sec.5 of the present
paper.

Adopting the line which we has taken, return now to the Hamilton
operators
\ref{HWEY},\ref{HRIV}. It is easy to see  that no linear combination
of them  can produce a scalar quantum potential.  However,  there is
a  combination $\hat  H^{(new)}_0$  which is consistent
with   Hamilton operator \ref{DW2}  in a particular
class  of coordinates  (and, consequently, of the position observables).

To show this, consider $\hat H^{\rm{(Weyl)}} _0 $ and
$\hat H^{\rm{(Riv)}}_0 $
in the normal quasi-Cartesian system of  coordinates
$\{y^a\}, $  with  its
origin at the point $ \xi^i$;  indices $\ a, b,...= 1,...,n $
are further  used  to denote components of  objects in
the coordinate system $\{y^a\} $.  In these coordinates, the values of
$y^a$ at an arbitrary point $ {\xi'}^i $  are  $y^a = h^a s  $
where $s$  is  the distance along the geodesic
connecting points $ \xi^i$ and $ {\xi'}^i $,  and  $h^a $  are
components of  the tangent vector to the geodesic with respect
an orthonormal $n$-tuple at the origin $ {\xi'}^i $.

In the normal quasi-Cartesian system,  the metric tensor $\omega_{ab} $,

its derivatives with respect to $y^a$, and, consequently, the
Christoffel
symbols ${\gamma^a}_{bc} (y) $ can be  expressed as a power series with
coefficients that are polynomials of components  of the curvature
tensor
$$
   {R^a}_{bcd} =\ptl_d {\gamma^a}_{bc} - \ptl_c {\gamma^a}_{bd} +
  {\gamma^a}_{de}{\gamma^e}_{bc} - {\gamma^a}_{ce} {\gamma^e}_{bd}
$$
and of the covariant derivatives of the tensor. Then, using the  Veblen
method of affine  extensions \cite{VEB} and the contracted Bianchi
identities, one  obtains the following
representations for   Hamilton operators:
\bea
{\hat H}^{\rm{(Weyl)}} _0 (y) & = & -\frac{\h^2}{2m} \left(\Delta(y)
+\frac14\left. R\right|_{y=0} + \frac14 \left.(\ptl_a R)\right|_{y=0}\
y^a +
O(y^2)\right).  \label{WE2}\\
 & = & -\frac{\h^2}{2m} \left(\Delta(\xi)
+\frac14 R(\xi) +  O(s^2)\right).  \label{WE3}\\
  {\hat H}^{\rm{(Riv)}}_0 (y)  & = & -\frac{\h^2}{2m}\left(\Delta(y)
+\frac13 \left.R\right|_{y=0} + \frac13 \left.(\ptl_a R)\right|_{y=0}\
y^a
+  O(y^2)\right). \label{RI2}\nonumber \\
& = & -\frac{\h^2}{2m} \left(\Delta(\xi)
+\frac13 R(\xi) +  O(s^2)\right).  \label{RI3}
\nda
(In transitions from eq.(\ref{WE2}) to eq.(\ref{WE3})
and from  eq.(\ref{RI2}) to  eq.(\ref{RI3}),  asymptotic relation
$$
R(\xi') = R(\xi) - \ptl_i R (\xi^i - \xi'^i)
= R(\xi) - \ptl_i R(xi) (\xi^i - \xi'^i) + O (s^2) = R(\xi) -
\ptl_a \left.R(y)\right|_{y=0}  h^a s  + O (s^2)
$$
is used).
It is seen from  here  that a Hamilton operator with  any coefficient of

$ R(\xi)$  can be obtained by appropriate  linear combination
of ${\hat H}^{\rm{(Weyl)}}_0 (y) $  and ${\hat H}^{\rm{(Riv)}}_0 $.
Our aim  is  to  get a {\it local} consistency between a certain
ordering rule in canonical quantization and  the WKB result
eq.(\ref{DW2}) in the
neighborhood of the origin $\xi'^i $,  We see easily
that    combination
\bea
{\hat H}^{\rm{(new)}} _0 (y)& = & 2 {\hat H}^{\rm{(Weyl)}} _0 (y) -
{\hat H}^{\rm{(Riv)}} _0 (y)  \label {TAG} \\
& = &  -\frac{\h^2}{2m}\left(\Delta(\xi)
+\frac16 R(\xi) +   O(s^2)\right) \nonumber
\nda
is just the Hamilton operator in the \Sche (\ref{DW2}) in
the  approximation indicated there.

  In terms of the basic operators
$ \hat p_j, \hat \xi^i$
\beq
{\hat H}^{\rm{(new)}} _0
= \frac1{2m} \hat p_i\ \omega^{ij}(\hat \xi)\ \hat p_j,
\nde
that is the  simplest of possible Hermitean  expressions which can be
constructed  from $\hat p_i, \hat p_j $ and $\omega^{kl} (\hat\xi)$.

The further logic  is  very simple. If an ordering rule is
universal for quantization of  any  $f (x,p) \in {\cal F}$
then one should consider the  combination of the Weyl and Rivier
orderings in the right-hand side of eq.(\ref{TAG}) as the  general
rule of ordering distinguished   by the condition of
local coincidence of WKB and  canonical QMs.  The corresponding
general formula of quantization  is the combination of quantizations
(\ref{WEY}) and (\ref{RIV})
\bea
f(p, \xi)\ &\qarr\ & (\hat f \psi) (\xi)     \nonumber \\
& = & (2 \pi \h)^{-n}\, \omega^{-\frac14} (\xi)\,
\int d^n\tilde\xi\, d^n p\
\exp\left(-\frac{\im}{\h} (\xi^i - {\tilde \xi}^i)p_i\right)
\Bigl(2f\bigl(p, \frac{\xi +\tilde \xi}{2}\bigr) \nonumber\\
& & -\ \frac{f(p, \xi)+ f(p, \tilde \xi)}{2}\Bigr)\
\omega^{\frac14}(\tilde\xi) \psi(\tilde \xi). \label{TAG2}
\nda

\section {The path  integral and  canonical formalism}

In  the path integral formalism  one usually starts with  the following
formal
expression for the propagator
\beq
{\Bbb K} (\xi'', t''|\xi', t')
= <\xi''| e^{-\frac\im\h (t''- t')\hat H_0}| \xi'>
\nde
which is proposed to be approximated as
\beq
{\Bbb K} (\xi'', t''|\xi', t') = \lim_{N\rightarrow\infty} \int \
\prod_{A=1}^{N-1} \sqrt{\omega (\xi_A)}\ d^n\xi_A \prod_{B=1}^{N-1}
<\xi_A|e^{-\frac\im\h \epsilon\hat H_0}|\xi_B>     \label{K}
\nde
where $\epsilon = (t''- t')/N$   and  $\xi_0 =\xi' , \ \xi_N =\xi''$.
The problem  is: from where does one  know  the Hamilton operator
$\hat H_0$ from?
At least, the following answers are possible: \\

1) To find it from  experiments.

2) To postulate it as a differential operator in
$L^2(V_n; {\Bbb C}; {\sqrt\omega} d^n \xi )$;
the standard postulate is $-(\h/2m^2) \Delta$.

3) To quantize canonically ${H_0}^{\rm{(cl)}}$, solving somehow the
problem of ambiguities.

4) To conjecture  the form of  ${\Bbb K} (\xi'', t''|\xi', t')$  for
$t'' \rightarrow t'$ and to determine $\hat H_0$  from the asymptotic
\Sche; this is just the  Pauli--DeWitt way \cite{DW}.

5) To deduce the \Sche as an asymptotic of the corresponding quantum
field
theory, as it is done  in  \cite{TAG1}. \\

Consider  approach 2) to the path integration following \cite{OLIV},
where the reader is referred for details.
These authors, as many others, consider the general covariance
of  $\hat H_0$  as a necessary condition. (From
my point of view, it is a sort of  prejudice that does not take into
attention the two-fold role of coordinates  mentioned above.)
Having taken the standard
expression $$-(\h/2m^2) \Delta$$ ,
one should express it in terms of the basic operators $\hat p_i,\ \hat
q^j$.
To this end,  one should to take any canonically obtained Hamilton
operator and to subtract the nonvanishing noncovariant quantum potential

$V_q$ corresponding to the ordering  chosen.
(In \cite{OLIV}, the latter is the Weyl ordering).  After that,
$ V_q $ will be inevitably present in the exponential  of  any
resulting form of the path integral. For example, it reads in the
lagrangean form \cite{OLIV}
\bea
{\Bbb K} (\xi'', t''|\xi', t') &=& \lim_{N\rightarrow\infty} \int \
\left(\frac{1}{2\pi \im\h\epsilon}\right)^{\pi N/2}
 \prod_{A=1}^{N-1} \sqrt{\omega (\xi_A)}\ d^n \xi_A \nonumber\\
&\times&\prod_{B=1}^{N-1}
\frac{\omega^{1/4}(\tilde
\xi_B)}{[\omega(\xi_B)\omega(\xi_{B-1})]^{1/4}}
\exp\left[\frac{\im}{\h} \epsilon
L^{\rm{(eff)}}\left(\tilde \xi_B,\ \frac{\xi_B
- \xi_{B-1}}{\epsilon}\right)\right]      \label{K2}
\nda
where  $ L^{\rm{(eff)}} (\xi, \dot \xi)
= L^{\rm{(cl)}}(\xi, \dot \xi) - V_q(\xi)$ and  $ \tilde \xi$ denotes
that the  value of the function $f$ taken in a special way
depending on the ordering chosen; if the  Weyl one  is
adopted, then
$$f(\tilde\xi_A) \std f\left(\frac{\xi_A  -\xi_{A-1}}2 \right))$$;
for the ordering introduced  in the present paper
$$ f(\tilde\xi_A) \std 2 f\left(\frac{\xi_A +\xi_{A-1}}{2}\right) -
\frac{f(\xi_A) + f(\xi_{A-1})}{2} $$

Thus, we come to the conclusion that the direct canonical approach,
that does not appeal to the general covariance,
leads to $L^{\rm{(eff)}} (\xi, \dot \xi) = L^{\rm{(cl)}}(\xi, \dot \xi)
$
in eq.(\ref{K2})  that corresponds better to the original Feynman idea.
However,  a dependence of the path integral  on the
choice of ordering remains in the form of  a rule  of   evalution
of $ f(\tilde\xi_A) $.

\section {Conclusion}
The approach adopted here can be considered as a sort of a
"physical" experiment with   different  mathematical
schemes of quantization.  A physical system (point-like particle
moving along a geodesic in $V_n$) taken as a probe  is the simplest
geometrically and physically meaningful one when
the ambiguities of quantization clearly manifest themselves.
At the same time, the system is sufficiently  fundamental
 that the results obtained by its consideration  might be used
to  select a preferred version  of the canonical quantization, in
general.
The system considered reveals also a  dependence of   the quantum
dynamics
on the  choice of the
basic observables (on "the quantum statics") .
In particular, this question seems to be  important for such fundamental

problem as quantization of gravitation where a separation of
the proper dynamics of the gravitational field from effects of the
choice of coordintes is one of the main dificulties.
In any case, our knowledge of the quantum theory  would not
be complete without clear realization the  basic questions
considered here.
 \\

{\bf Acknowledgement}\\

The author is deeply grateful to the Russian Foundation for
Basic Research which supported this work  by Grant No 00-01-00871
and to Professor B.M.Barbashov  for a useful discussion.

\end{document}